\documentclass[twocolumn,aps,prd]{revtex4-2}
\usepackage{graphicx}
\usepackage{bm}
\usepackage{relsize}
\usepackage{xfrac}
\begin{document}
\newcommand*{\bi}{\bibitem}
\newcommand*{\ea}{\textit{et al.}}
\newcommand*{\eg}{\textit{e.g.}}
\newcommand*{\plb}[3]{Phys.~Lett.~B \textbf{#1}, #2 (#3)}
\newcommand*{\phrc}[3]{Phys.~Rev.~C~\textbf{#1}, #2 (#3)}
\newcommand*{\phrd}[3]{Phys.~Rev.~D~\textbf{#1}, #2 (#3)}
\newcommand*{\phrl}[3]{Phys.~Rev.~Lett.~\textbf{#1}, #2 (#3)}
\newcommand*{\ptep}[3]{Prog. Theor. Exp. Phys. \textbf{#1}, #2 (#3)}
\newcommand*{\ibid}[3]{\textit{ ibid.} \textbf{#1}, #2 (#3)}
\newcommand*{\ra}{\rightarrow}
\newcommand*{\kpkm}{K^+K^-}
\newcommand*{\knkn}{K^0\bar{K^0}}
\newcommand*{\kskl}{K^0_SK^0_L}
\newcommand*{\rf}[1]{(\ref{#1})}
\newcommand*{\be}{\begin{equation}}
\newcommand*{\ee}{\end{equation}}
\newcommand*{\die}{e^+e^-}
\newcommand*{\jj}{\mathrm i}
\newcommand*{\ndf}{\mathrm{NDF}}
\newcommand*{\cndf}{\chi^2/\mathrm{NDF}}
\newcommand*{\minuit}{\texttt{MINUIT}~}
\newcommand*{\ie}{\textit{i.e.}}
\newcommand*{\dek}[1]{\times10^{#1}}
\def\babar{\mbox{\slshape B\kern-0.1em{\smaller A}\kern-0.1em
    B\kern-0.1em{\smaller A\kern-0.2em R}}}

\title{New narrow resonance in the $\bm{e^+e^-\rightarrow \phi\,\eta}$ data by
Belle collaboration}
\author{Peter Lichard}
\affiliation{Institute of Physics and Research Centre for Computational 
Physics and Data Processing, Silesian University in Opava, 746 01 Opava, 
Czech Republic\\
and\\
Institute of Experimental and Applied Physics,
Czech Technical University in Prague, 128 00 Prague, Czech Republic}

\begin{abstract}
Fitting the recent $e^+e^-\ra\phi\,\eta$ data by the Belle collaboration 
with a theoretical formula reveals, besides the dominant $\phi(1680)$ 
resonance, two narrow resonances: expected $\phi(2170)$ resonance and an 
unexpected resonance with the mass of about 1851 MeV. Close proximity to 
the $X(1835)$ resonance suggests that the new resonance may be interpreted 
as the $p \bar p$ baryonium in an excited state.  Follow-up analysis found 
the same resonance also in $e^+e^-\ra\omega\,\eta$ data by CMD-3 experiment.
\end{abstract}
\maketitle
Recently, a study of the $e^+e^- \ra \eta\,\phi$ process with the Belle 
detector at the KEKB asymmetric-energy $\die$ collider has been published 
\cite{belle2023}. Experimentalists explored the Initial State Radiation 
method and covered the $\die$ invariant energy range from 1.56 to 3.96 GeV 
in 120 bins. The published values of the $e^+e^- \ra \eta\,\phi$ cross 
section are accompanied by statistical and systematic errors.

As the members of the Belle Collaboration stated in the Introduction, one of 
the experiment's goals was to study the properties of the $\phi(2170)$ 
resonance. This resonance was discovered in 2006 by the \babar~Collaboration 
at the Stanford Linear Accelerator Center in $\die\ra\phi\,f_0(980)$ reaction 
\cite{babar2006} and later confirmed by several experiments in various 
processes. Of those, we list two that confirmed the $\phi(2170)$ resonance 
in the $\die$ annihilation into the $\eta(547)\phi(1020)$ system: 
\babar~\cite{babar2008} and BESIII experiment \cite{besiii2021} at the 
Beijing Electron Positron Collider.

When analyzing their cross-section data, the Belle collaboration first fit 
them by assuming one resonance. They got the parameters of the dominant 
$\phi(1680)$ resonance correctly, see Table 1 in \cite{belle2023}, even if 
the quality of the fit was not excellent [$\cndf=85/60$, which translates 
to Confidence Level (CL) of 2\%]. Then, they used a phenomenological fitting 
procedure tailored for two resonances to find the signs of the $\phi(2170)$ 
resonance. Again, the parameters of the dominant $\phi(1680)$ resonance were 
varied, whereas those of the other resonance were fixed at the values obtained 
for $\phi(2170)$ by BESIII Collaboration \cite{besiii2021}. No significant 
$\phi(2170)$ signal was found.

To investigate the reason for this conundrum, we decided to perform our own 
analysis of the Belle \cite{belle2023} cross-section data based on a
theoretical formula capable of handling, in principle, any number of
resonances.

\begin{figure}[b]
\includegraphics[width=0.39\textwidth,height=0.1\textwidth]{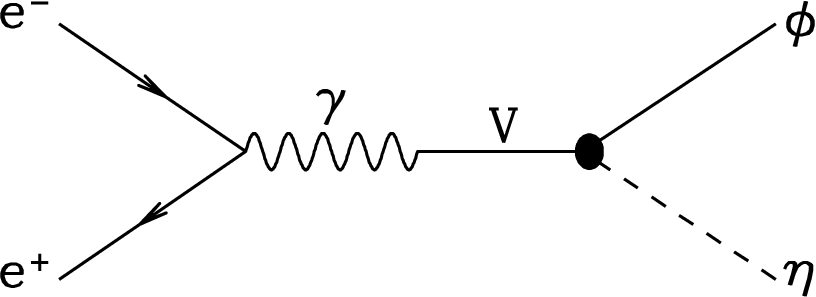}
\caption{\label{fig:ee2phieta}Feynman diagram defining our VMD model}
\end{figure}

For the description of the electron-positron annihilation into the vector
meson $\phi$ and pseudoscalar meson $\eta$ we use a Vector Meson
Dominance (VMD) model based on the Feynman diagram depicted in 
Fig.~\ref{fig:ee2phieta} and the interaction Lagrangian
\[
{\cal L}_{V\!\phi\eta}(x)=\frac{{g_{}}_{V\!\phi\eta}}{m_{V}}
\epsilon_{\mu\nu\rho\sigma}\partial^\mu V^\nu(x)\,\partial^\rho
\phi^\sigma(x)\,\eta(x)\,,
\]
where particle symbols denote the corresponding quantum fields.
The $\gamma V$ junction is parametrized as $eM_V^2/g_V$ in analogy with
the $\gamma\rho^0$ junction $eM^2_{\rho^0}/g_\rho$. 
Further, we define dimensionless 
quantity $r=g_{V\!\phi\eta}/g_{V}$. When we consider several 
intermediate vector mesons $V_i$, the $e^+\ e^-\ra \phi\,\eta$ cross section 
comes out as
\be
\label{xsect}
\sigma=\frac{\pi\alpha^2}{6}\frac{\lambda^{\sfrac{3}{2}}(s)(s+2z)}
{s^3\sqrt{s(s-4z)}}
\left|\sum_{i=1}^n\frac{r_iM_ie^{\jj\delta_i}}{s-M_i^2+\jj 
M_i\Gamma_i}\right|^2,
\ee
where $x=m_\phi^2$, $y=m_\eta^2$, $z=m_e^2$,
$\lambda(s)=s^2+x^2+y^2-2sx-2sy-2xy$, and $\delta_1=0$.

Performing the fits with one or two resonances, we got similar 
results as Belle Collaboration. One-resonance fit yielded a resonance with 
parameters close to those listed by the Particle Data Group \cite{pdg2022} 
for the $\phi(1680)$
\cite{omega1650}.

When doing the fit with two resonances, we did not fix the mass and width of 
one of them to the expected $\phi(2170)$ values, which the Belle collaboration 
did. Even thus, we got a clear signal of only one resonance, namely 
$\phi(1680)$. The parameters of the second one do not correspond to any 
conceivable resonance. They reflect the effort of the minimalization 
program \cite{fred} to bring the theoretical curve closer to the data in the 
vast region around 1920 MeV. 

\begin{table}[t]
\caption{\label{tab:3reso}Parameters of the fits to the Belle data 
\cite{belle2023} up to 3 GeV based on Eq. \rf{xsect}. The statistical
significance of the $i$th resonance is denoted as $\Sigma_i$.}
\begin{tabular*}{8.5cm}[b]{lccc}
\hline
\hline
~~~~~~~~~~~~~~~~~&~~1 resonance~~&~~2 resonances~~&~~~3 resonances~~~\\
\hline
$r_1$           & 0.3761(94)     & 0.291(29)      &  0.360(14)    \\
$M_1$ (MeV)     & 1650.5$\pm$4.1 & 1661.8$\pm$6.0 & 1656.8$\pm$4.9  \\
$\Gamma_1$ (MeV)& 158.7$\pm$5.3  & 125$\pm$12     & 150.8$\pm$7.0   \\
$\Sigma_1$      & $40\,\sigma$   & $10\,\sigma$   & $25\,\sigma$     \\  
$r_2$           &                & 0.050(32)      & 0.0077(43)   \\
$M_2$ (MeV)     &                & 1921$\pm$86    & 1850.7$\pm$5.3  \\
$\Gamma_2$ (MeV)&                & 290$\pm$230    & 25$\pm$35    \\
$\delta_2$      &                & 0.8$\pm$1.2    & 5.59(44)     \\
$\Sigma_2$      &                & $1.5\,\sigma$  & $1.7\,\sigma$  \\   
$r_3$           &                &                & 0.0044(22)    \\
$M_3$ (MeV)     &                &                & 2215.7$\pm$8.3 \\
$\Gamma_3$ (MeV)&                &                & 35$\pm$23 \\
$\delta_3$      &                &                & 2.59(39)    \\
$\Sigma_3$      &                &                & $2.0\,\sigma$ \\
\hline
$\chi^2$/NDF    & 83.6/69       & 58.5/65        & 47.1/61         \\
CL (\%)         & 11.1          & 70.2           & 90.4         \\
\hline
\hline
\end{tabular*}
\end{table}

\begin{figure}[h]
\includegraphics[width=8.6cm]{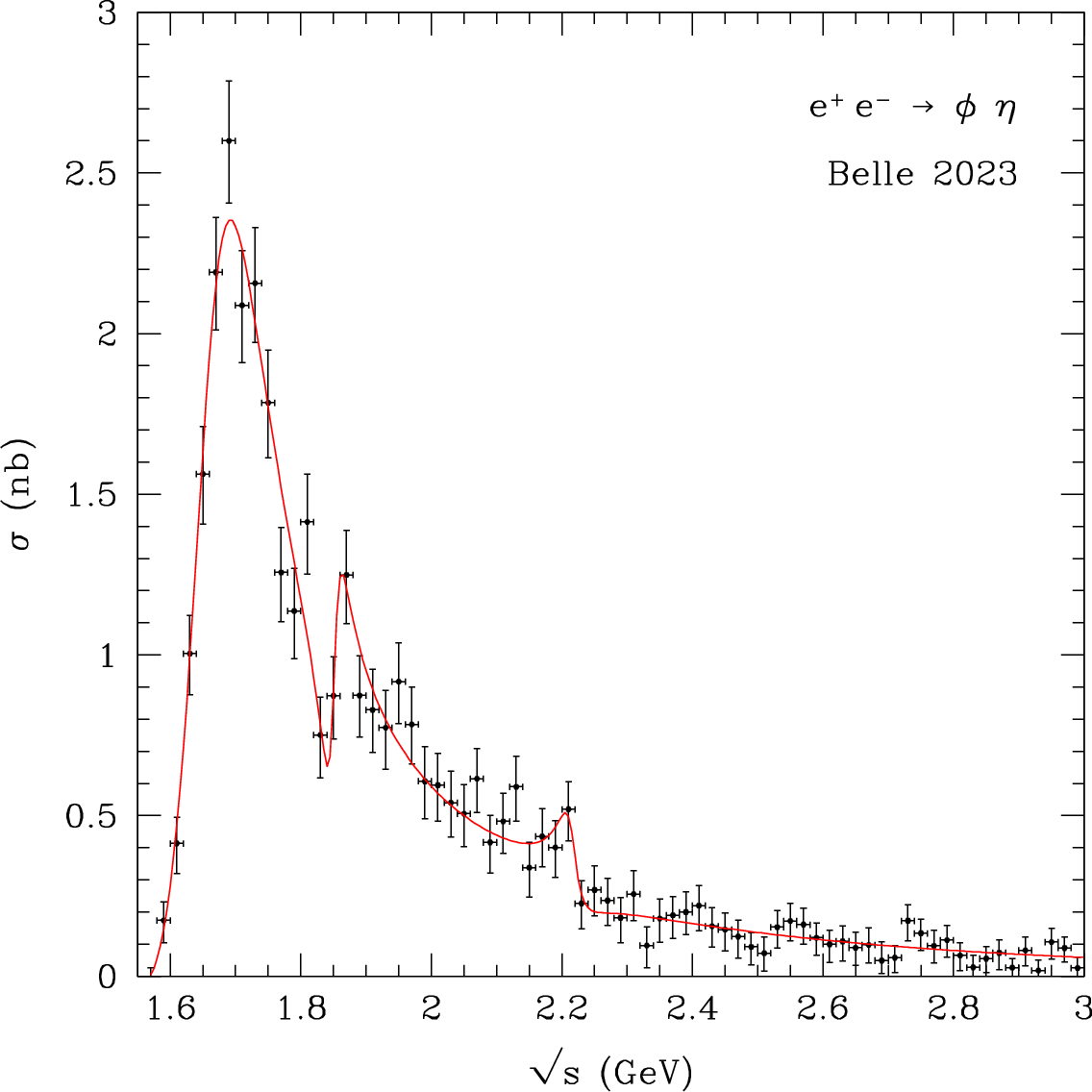}
\caption{\label{fig:3reso}The excitation curve obtained by the Belle 
collaboration \cite{belle2023}  and our fit to it using formula \rf{xsect} 
with three resonances. Only statistical errors of data are shown
and were used in fitting. The parameters of the fit are provided in 
Table~\ref{tab:3reso}.}
\end{figure}
However, when we allowed three resonances, the situation drastically changed. 
The quality of the fit increased to CL=90.4\%, and two narrow resonances 
appeared accompanying the dominant $\phi(1680)$ resonance; see Fig. 
\ref{fig:3reso} and Table \ref{tab:3reso}. The one with the higher mass lies 
in the region where the $\phi(2170)$ resonance is expected. The mass of it is 
higher than the PDG average but agrees with the three BESIII measurements 
\cite{bes3}. Here, $\phi(2170)$ manifests as a sudden drop of the excitation
curve, not as a peak in some experiments. The width we found is smaller than 
the PDG average. However, it agrees with those obtained in several experiments 
listed in \cite{pdg2022}.

The statistical significance of the newly found resonance with a mass of 
$(1850.7\pm5.3)$~MeV and width of $(25\pm35)$~MeV is low. 
There is a possibility that it is not a true resonance but a mere product of 
statistical fluctuation in data.
To investigate this issue, we use 
the following ``look everywhere'' method: The minimalization of the $\chi^2$ 
procedure is repeated many times with starting values of resonances 1 and 3 
kept at values from Table \ref{tab:3reso}. The starting value of $M_2$ is 
randomly generated in the interval (1600, 2900)~MeV, that of $\Gamma_2$ in the 
interval (10, 40)~MeV. The other starting values are chosen at $r_2=0$ and 
$\delta_2=\pi$. After the minimalization procedure, the observed new 
``resonances'' were grouped into clusters with the masses within a 
narrow interval (we chose a width of 12 MeV). After repeating the procedure 
a thousand times, we identify 20 clusters (some with only a few entries), 
of which the most populated are shown in Table \ref{tab:everywhere}.
Judging from the number of entries in clusters and the mean values of 
$\chi^2$, the behavior of the excitation curve around 1851 MeV satisfies 
the resonance requirement better than other parts of the spectrum outside 
the two established resonances. Also, the extremely narrow widths of the other 
``resonances'' shown in Table \ref{tab:everywhere} indicate that they are 
products of statistical fluctuations. All this makes the resonance 2 in the 
rightmost column of Table \ref{tab:3reso} the only plausible candidate for the 
true resonance accompanying $\phi(1680)$ and $\phi(2170)$ in the Belle data.
Of course, the statistical fluctuation origin of a resonance there cannot 
be completely ruled out.

\begin{table}[b]
\caption{\label{tab:everywhere}Mean mass, width, and $\chi^2$ together with
number of resonances in the most populated clusters after a thousand
randomly generated searches.}
\begin{tabular*}{5cm}[b]{cccc}  
\hline
\hline
$\overline M$~(MeV)      &$\overline\Gamma$~(MeV) &$\overline{\chi^2}$&~~~n\\
\hline
1850.8           & 21.7 &  47.2 &~~~247\\
2734.5           & 0.9  &  54.3 &~~~117 \\
2529.4           & 1.9  &  55.2 &~~~106 \\
2396.2           & 5.7  &  58.4 &~~~~~77 \\
\hline
\hline
\end{tabular*}
\end{table}

When we accept the possibility that the new resonance is a real effect, 
we should think about its origin. The mass of 
(1850.7$\pm$5.3)~MeV and width of (25$\pm$35)~MeV points toward the $X(1835)$ 
resonance. However,  the quantum numbers $J^{PC}=0^{-+}$ of the latter prevent 
it from being produced in the direct channel of the $\die$ annihilation,
which requires $J^{PC}=1^{--}$. In the 
listing dealing with $X(1835)$, the PDG \cite{pdg2022} mentions the possibility
that this object is a superposition of two states differing in widths. More
specifically, the results of BESIII experiment \cite{besiiitwo} ``suggest the 
existence of either a broad state around 1.85~GeV/$c^2$ with strong coupling 
to the $p\bar p$ final states or a narrow state just below the $p\bar p$ mass 
threshold''. The latter's existence was proposed long ago \cite{boundstate}
as a $p\bar p$ state bound by strong interactions (hereafter, we call it 
protonium). The idea was later elaborated in several papers.

The quantum numbers of the $X(1835)$ suggest that its protonium component is 
in the $L\!=\!S\!=\!J\!=\!0$ state. Because of the large bounding energy (BE) 
\cite{pdgbind}
we may expect protonium's excited states to exist below the $2m_p$ threshold. 
Those of them with quantum numbers $L=0,\, S\!=\!J\!=\!1$  or  $L\!=\!2, 
S\!=\!J\!=\!1$ provide protonium with $J^{PC}=1^{--}$, which can appear in 
the intermediate state of $\die$ annihilation. Guided by this, we suggest 
that the new narrow resonance in the Belle data \cite{belle2023} is an 
excited state of the protonium. 

The situation is similar to strongly bound kaoniums ($\kpkm$) and ($\knkn$), 
the ground states of which cannot be produced in the direct channel of $\die$ 
annihilation. Their excited states with $L\!=\!1$ were detected as subthreshold 
poles in the  $\die\ra\kpkm$ and $\die\ra\kskl$ processes, respectively 
\cite{kaonium}. 

The BE of the excited protonium calculated from its mass from Tab.
\ref{tab:3reso} comes out as 26~MeV. For comparison, let us recall that the 
BE of excited kaoniums was estimated at 10~MeV \cite{kaonium}. Salnikov and
Milstein \cite{salnikov} have recently predicted a bound state of $\Lambda_c$ 
and its antiparticle with BE of 38~MeV.

\begin{figure}[t]
\includegraphics[width=8.6cm]{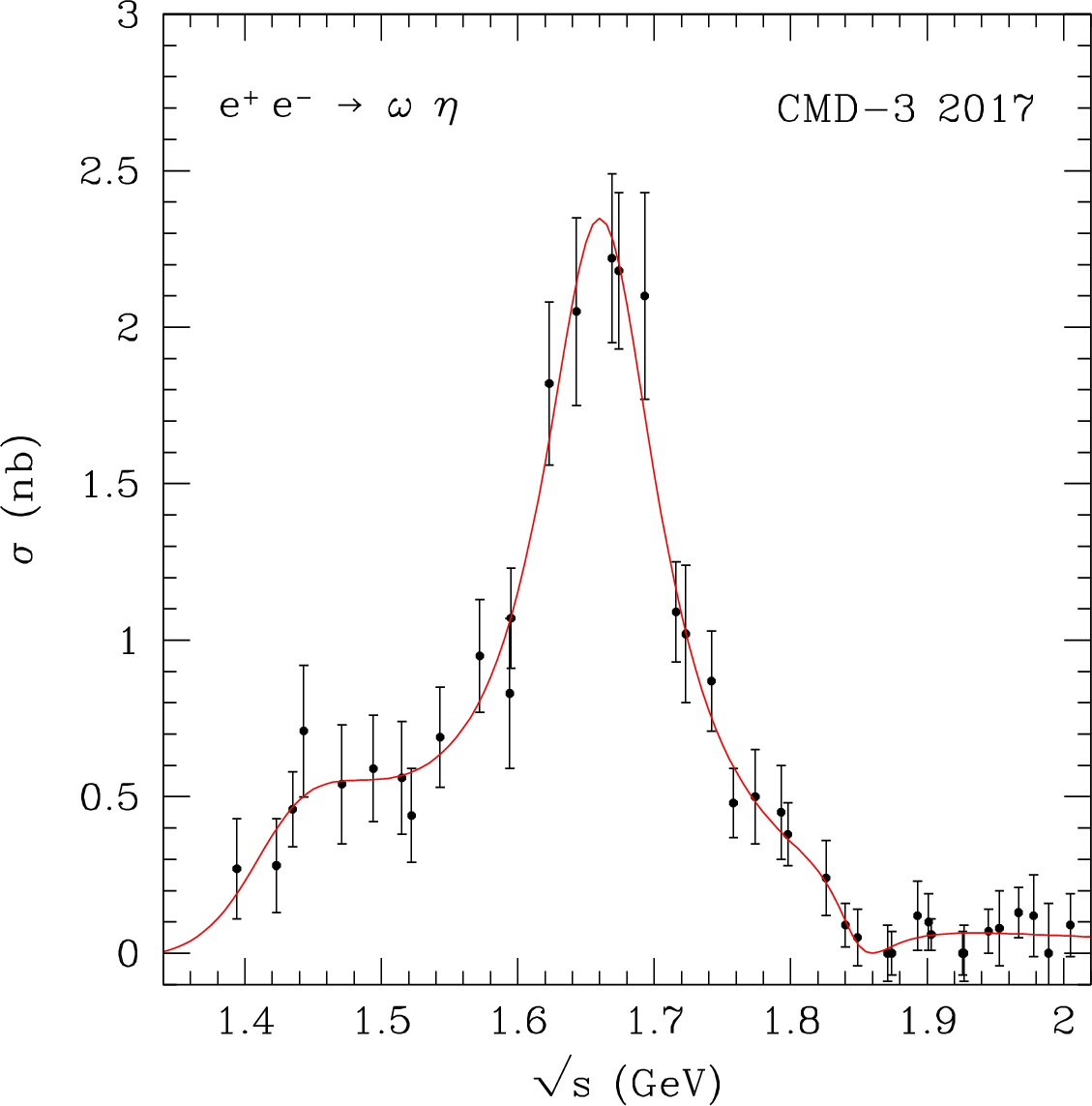}
\caption{\label{fig:cmd3reso}3-resonance fit to the cross section data measured 
in the  CMD-3 experiment~\cite{cmd32017} using formula \rf{xsect}. The 
parameters of the fit are given in Table~\ref{tab:cmd3}.}
\end{figure}

We have started scanning other sets of the $\die$ annihilation data. Up to now, 
we have found indication of excited protonium in the $\die\ra\omega\,\eta$ data 
of the CMD-3 experiment \cite{cmd32017} at Budker Institute of Nuclear Physics 
in Novosibirsk, see Figure \ref{fig:cmd3reso} and Table \ref{tab:cmd3}.
\begin{table}[h]
\caption{\label{tab:cmd3}Parameters of the 3-resonance fit to the CMD-3 data 
\cite{cmd32017} based on Eq. \rf{xsect}.  The statistical significance 
of the $i$th resonance is denoted as $\Sigma_i$.}
\begin{tabular*}{8.6cm}[b]{lccc}
\hline
\hline
$i$           & 1 & 2 & 3\\
\hline
$r_i$           & 0.102(97)   & 0.092(38)       &  0.013(18) \\
$M_i$ (MeV)&~~~~~~1420$\pm$60~~~~~~&~~~1660.0$\pm$8.4~~~&~~~~~1847$\pm$16~~~~~\\
$\Gamma_i$ (MeV)& 136$\pm$115 & 106$\pm$15      & 52$\pm$31\\
$\delta_i$      & 0           & 1.79(59)        & 5.3$\pm$1.0\\
$\Sigma_i$    & 1.0\,$\sigma$ &2.4\,$\sigma$&  0.7\,$\sigma$ \\
\hline
\multicolumn{4}{c}
 {$\chi^2$/NDF=11.8/29~~~~~~~~~~Confidence level = 99.8\%}         \\
\hline
\hline
\end{tabular*}
\end{table}
The excited protonium mass and width (1847$\pm$16~MeV, 52$\pm$31~MeV) agree 
with those from Belle data \cite{belle2023} (1850.7$\pm$5.3~MeV,  
25$\pm$35~MeV).  Unfortunately, its statistical significance is only 
0.7\,$\sigma$. 

\textbf{To conclude:} In this work, we indicated the possible existence of a resonance 
with the mass 
and width resembling that of the X(1835) resonance but with different quantum 
numbers $J^{PC}=1^{--}$. It may be interpreted as an excited state of the 
protonium, a strongly bound $p\bar p$ system, widely considered one of two 
components of the X(1835) resonance. Unfortunately, low statistical 
significance does not allow claiming the new resonance's evidence (3\,$\sigma$). 
Additional confirmation is needed by analyzing existing data or by a new 
measurement. 

\begin{acknowledgments}
I thank Filip Blaschke, Josef Jur\'{a}\v{n}, and Santu Mondal for the useful 
discussions.
\end{acknowledgments}

\end{document}